\RequirePackage{fix-cm}
\documentclass[twocolumn, epjc3]{svjour3}  
\smartqed  
\RequirePackage{graphicx}
\RequirePackage{epsfig}
\RequirePackage{lineno}
\RequirePackage{subfigure}
\journalname{Eur. Phys. J. C}
\begin{document}

\title{Enriched Zn$^{100}$MoO$_4$ scintillating bolometers to search for $0 \nu 2\beta$ decay of $^{100}$Mo with the LUMINEU experiment
}


\author{A.S. Barabash\thanksref{addr0}
        \and
	D.M.~Chernyak\thanksref{addr1,addr2}
	\and
        F.A.~Danevich\thanksref{addr1}
	\and
	A.~Giuliani\thanksref{e1,addr2,addr3,addr4}
     	\and
	I.M. Ivanov\thanksref{addr5}
	\and
	E.P. Makarov\thanksref{addr5}
	\and
     	M. Mancuso\thanksref{addr2,addr3} 
	\and
	S. Marnieros\thanksref{addr2}
     	\and
	S.G. Nasonov\thanksref{addr5}
	\and
        C. Nones\thanksref{addr6}
        \and 
     	E.~Olivieri\thanksref{addr2}
     	\and
	G. Pessina\thanksref{addr4}
	\and
	D.V. Poda\thanksref{addr1,addr2}
	\and
	V.N. Shlegel\thanksref{addr5}
	\and
	M.~Tenconi\thanksref{addr2}
	\and
     	V.I.~Tretyak\thanksref{addr1,addr6b}
	\and
	Ya.V. Vasiliev\thanksref{addr5} 
	\and
	M. Velazquez\thanksref{addr7}
	\and
	V.N. Zhdankov\thanksref{addr8} 
}

\thankstext{e1}{e-mail: andrea.giuliani@csnsm.in2p3.fr}


\institute{Institute of Theoretical and Experimental Physics, 117218 Moscow, Russia\label{addr0}
\and
Institute for Nuclear Research, MSP 03680 Kyiv, Ukraine\label{addr1}
\and
Centre de Sciences Nucl\'eaires et de Sciences de la Mati\`ere, 91405 Orsay, France\label{addr2}
\and
Dipartimento di Scienza e Alta Tecnologia dell'Universit\`a dell'Insubria, 22100 Como, Italy\label{addr3}
\and
INFN, Sezione di Milano Bicocca, 20126 Milano, Italy\label{addr4}
\and
Nikolaev Institute of Inorganic Chemistry, 630090 Novosibirsk, Russia\label{addr5}
\and
Service de Physique des Particules, CEA-Saclay, 91191 Gif sur Yvette, France\label{addr6}
\and
INFN, Sezione di Roma ``La Sapienza'', 00185 Rome, Italy\label{addr6b}
\and
CNRS, Universit\'e de Bordeaux, ICMCB, 33608 Pessac, France\label{addr7}
\and
CML Ltd., 630090 Novosibirsk, Russia\label{addr8}
}


\maketitle

\begin{abstract}
The LUMINEU project aims at performing a demonstrator underground experiment searching for the neutrinoless double beta decay of the isotope $^{100}$Mo embedded in zinc molybdate (ZnMoO$_4$) scintillating bolometers. In this context, a zinc molybdate crystal boule enriched in $^{100}$Mo to 99.5\% with a mass of 171 g was grown for the first time by the low-thermal-gradient Czochralski technique. The production cycle provided a high yield (the crystal boule mass was 84\% of the initial charge) and an acceptable level -- around 4\% -- of irrecoverable losses of the costly enriched material. Two crystals of 59 g and 63 g, obtained from the enriched boule, were tested aboveground at milli-Kelvin temperature as scintillating bolometers. They showed a high detection performance, equivalent to that of previously developed natural ZnMoO$_4$ detectors. These results pave the way to future sensitive searches based on the LUMINEU technology, capable to approach and explore the inverted hierarchy region of the neutrino mass pattern.

\end{abstract}

\section{Introduction}

Neutrinoless double beta ($0\nu2\beta$) decay is a crucial process in particle physics since it provides the only experimentally viable method to ascertain the Majorana nature of neutrino, performing in the meantime a sensitive test of the lepton number conservation. In addition, it has the potential to establish the absolute neutrino mass scale and to give information about the hierarchy of the neutrino masses~\cite{Cre14,Sch13,Ell12,Ver12,Giu12,Gom12,Rod11,2bTables}. In particular, one of the main objectives of next-generation $0 \nu 2 \beta$ decay experiments is to explore the inverted hierarchy region of the neutrino mass pattern. This implies reaching a lifetime sensitivity of the order of $10^{26}-10^{27}$ years, which requires a background level close to zero in the ton$\times$year exposure range.  

A powerful technology to achieve this outstanding performance is represented by scintillating bolometers, used to study isotopes with a $0\nu2\beta$ Q-value definitely higher than 2.6~MeV, and therefore free from the $\gamma$ background induced by natural radioactivity. In these devices, the dominant background is expected to be given by energy-degraded $\alpha$ particles, which however can be rejected thanks to the different scintillation-to-heat ratio between $\alpha$ and $\beta$ particles~\cite{giusang,Mil-scint,Pirr06}.

The LUMINEU project~\cite{Lumin} (Luminescent Underground Molybdenum Investigation for NEUtrino mass  and nature) aims at developing scintillating bolometers based on zinc molybdate (ZnMoO$_4$) crystals to study the isotope $^{100}$Mo (Q-value=3034~keV \cite{Rah08}), capable reaching the performance required to explore the inverted hierarchy region, as shown for the first time in Ref.~\cite{Bee12} and susbsequently confirmed in Ref.~\cite{cuore-luce}. A crytical step in the LUMINEU path is represented by the growth of high-quality radio-pure large-mass (300 -- 500 g) ZnMoO$_4$ monocrystals. As far as non-isotopically enriched material is concerned, the required crystallization technology for a sensitive $0 \nu 2 \beta$ decay experiment has been established, and individual detectors -- prefiguring the single modules of future large arrays -- have already shown to be able to reach the desired bolometric performance and intrinsic radiopurity levels~\cite{Bee12,Bee12a,Bee12b,Bee12c,Che13,Ber14}.  

However, the crucial step of reproducing these excellent preliminary results with enriched molybdenum (highly desirable operation for a future large-scale experiment, the $^{100}$Mo  natural abundance being only 9.7\% \cite{Ber11}) is not trivial for two main reasons. On one hand, the initial chemical and radioactive purity levels of enriched samples, normally poorer than those characterizing natural material, may conflict with the construction of well performing bolometric detectors. This problem was observed for instance in TeO$_2$ bolometers~\cite{Arn08}. On the other hand, enriched material is very expensive and its procurement represents the highest cost factor in future searches. Therefore, the purification / crystallization chain must imply negligible irrecoverable losses of the $0\nu2\beta$ decay candidate isotope. This strong requirement is not a priori compatible with all the purification and crystal-growth technologies. In this letter, we show for the first time that both potential obstacles are actually overcome in the ZnMoO$_4$ case, adding a further essential element in favor of the use of the LUMINEU technology for future $0\nu2\beta$ decay searches.

\section{Production of a Zn$^{100}$MoO$_4$ crystal boule}
\label{sec:cryst}

Molybdenum enriched in the isotope $^{100}$Mo up to 99.5\% was used to
develop zinc molybdate crystals. The enriched molybdenum was produced at the Kurchatov Institute in the eighties of the last century (former Soviet Union). Approximately one kilogram of the material in form of metal, belonging to the Insitute for Nuclear Research (Kyiv, Ukraine) and now available for the LUMINEU program, was utilized in an experiment at the Modane underground laboratory (France) to search for double beta decay of $^{100}$Mo to excited states of $^{100}$Ru \cite{Blum92}. Afterwards, in order to improve its purity level, the
metallic sample of enriched $^{100}$Mo was dissolved in 20\%
ultrapure nitric acid and transformed into molybdenum acid
($^{100}$MoO$_3 \cdot n$H$_2$O). After rinsing the compound by
nitric acid solution and annealing, a sample of 1199 g of purified molybdenum
oxide ($^{100}$MoO$_3$) was obtained and used in the ARMONIA
experiment \cite{ARMONIA}. The purification procedure has
effectively removed the pollution in $^{40}$K and $^{137}$Cs by
one order of magnitude. The concentrations of thorium and radium
were also decreased by factors 2 and 4, respectively. The
following radioactive contamination of the $^{100}$MoO$_3$ sample
can be derived from the data of the experiment (in
mBq/kg): 2.0(2) for $^{226}$Ra ($^{238}$U chain), 0.8(1) for
$^{228}$Th ($^{232}$Th chain), 5.9(1) for $^{137}$Cs, and 36(2) for
$^{40}$K  \cite{ARMONIA} .

The level of impurities in the $^{100}$MoO$_3$ was also measured
by inductively coupled plasma mass-spectrometry (ICP-MS) and
atomic absorption spectroscopy (AAS) methods. The results of these
measurements, presented in Table \ref{100Mo-cont}, as well as
radioactive contamination of the sample, suggest that the material
should be additionally purified to be used for crystal
scintillator production. For example, the concentration of the isotope $^{228}$Th in ZnMoO$_4$ crystals to be used in future $0 \nu 2 \beta$ decay searches must not be higher than 10~$\mu$Bq/kg in order to approach the requested zero background conditions mentioned in the Introduction~\cite{Bee12}. 

\begin{table}[htb]
\caption{Contamination of $^{100}$MoO$_3$ measured by inductively
coupled plasma mass-spectrometry (ICP-MS) and atomic absorption
spectroscopy (AAS) methods.}
\begin{center}
\begin{tabular}{p{2cm} p{2cm} p{2cm}}
\hline
\    &  \multicolumn{2}{c} {Concentration of element} \\
Element & \multicolumn{2}{c} {in $^{100}$MoO$_3$ (ppm)} \\
\cline {2-3}
 ~          & ICP-MS            & AAS \\
 \hline
  Na        & --                & $<60$  \\
  Mg        & $<0.5$            & $<4$              \\
  Al        & 2.4               & --              \\
  Si        & --                & $<500$       \\
  K         & $<15$             & $<10$              \\
  Ca        & --                & $<10$    \\
  V         & 0.05              & --                 \\
  Cr        & 0.2               & $<5$                 \\
 Mn         & 0.1               & --                 \\
 Fe         & 8                 & $<5$                \\
 Ni         & 0.01              & --                 \\
 Cu         & 0.1               & --                \\
 Zn         & 0.1               & $<4$                 \\
 Ag         & 0.3               & --                 \\
 W          & 1700              & 550                \\
 Pb         & 0.008             & --                 \\
 Th         & $<0.0005$         & --         \\
 U          & 0.001             & --         \\
 \hline
\label{100Mo-cont}
\end{tabular}
\end{center}
\end{table}

A two-stage technique of molybdenum purification, consisting of
sublimation of molybdenum oxide in vacuum and recrystallization
from aqueous solutions by co-precipitation of impurities on zinc
molybdate sediment, was applied to purify the enriched molybdenum.
The purification procedure is reported in details in
Ref.~\cite{Ber14}.

Sublimation was carried out with the addition of zinc oxide to
reduce the concentration of tungsten:

\begin{center}
 ZnMoO$_4$ + WO$_3$ $\to$ ZnWO$_4$ + MoO$_3$ $\uparrow$.
\end{center}

\noindent The residuals after the sublimation process
were analyzed at the analytical laboratory of the Nikolaev
Institute of Inorganic Chemistry with the help of atomic
emission spectrometry. They contain Fe (0.05 wt\%), Si
($0.6-0.8$ wt\%), Mo ($11-12$ wt\%), W ($22-33$ wt\%) and Zn ($12-16$
wt\%). It should be stressed that the analysis confirmed a contamination
by iron in the initial MoO$_3$. The sublimates were then
annealed in air atmosphere to obtain yellow color stoichiometric
MoO$_3$. Losses of enriched molybdenum at this stage of
purification did not exceed 1.4\%.

After the sublimation process, the molybdenum oxide contained
needlelike granules up to a few millimeter long. The presence of these
granules (which cause difficulties to synthesize zinc
molybdenum compound) is one more argument to apply an additional
stage of purification, consisting of recrystallization of the obtained
molybdenum oxide in aqueous solution. To this purpose, the
molybdenum oxide was dissolved in ammonia solution at room
temperature using zinc molybdate as a collector:

\begin{center}
MoO$_3$ + 2NH$_4$OH = (NH$_4$)$_2$MoO$_4$ + H$_2$O.
\end{center}

The molybdenum losses during recrystallization from
aqueous solutions are expected to be recovered in conditions
of mass production. Therefore, we have estimated the losses at
this stage taking into account a purification process of about 20 kg
of natural molybdenum performed in the framework of the LUMINEU
program and its follow-up. The losses do not exceed 2\%.

Powder of Zn$^{100}$MoO$_4$ (203.98 g) was obtained by solid-phase
synthesis of high-purity zinc oxide (72.23 g, produced by Umicore)
and purified enriched molybdenum oxide $^{100}$MoO$_3$ (131.75 g). The mixture of the two compounds was kept at a temperature of $\approx 680$~$^{\circ}$C in a platinum cup over 12 hours.

A zinc molybdate crystal boule from the enriched $^{100}$Mo compound was grown by the low-thermal-gradient Czo\-ch\-ral\-ski technique
\cite{Pavl92,Boro01,Gala09} in a platinum crucible with size $\oslash 40\times
100$ mm. The crystal was grown at a rotation speed of 20 rotations per
minute at the beginning of the process, decreasing to 4 rotations
per minute at the end. The temperature gradient did not exceed 1
$^{\circ}$C/cm.

\begin{figure}[htbp]
\begin{center}
\mbox{\epsfig{figure=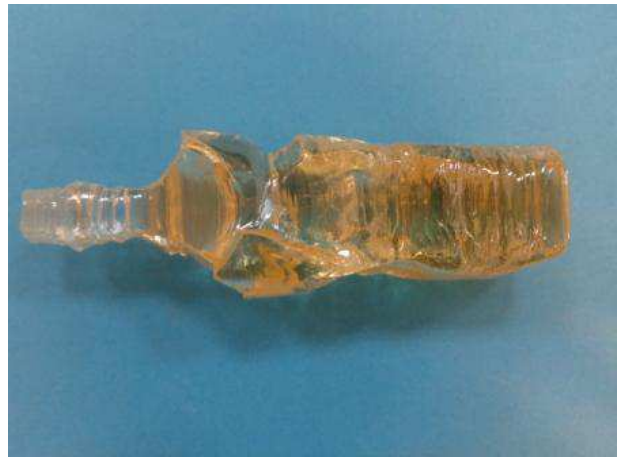,height=6.0cm}} \caption{Boule
of Zn$^{100}$MoO$_4$ single crystal with mass of 170.7 g and
length of 95 mm grown by the low-thermal-gradient Czochralski
process.}
 \label{fig:boule}
\end{center}
\end{figure}

The mass of the crystal boule, shown in Fig.~\ref{fig:boule}, is of 170.7~g. Some coloration of the crystal (in contradiction with the practically colorless samples produced from natural molybdenum~\cite{Che13}) can be explained by remaining traces of iron in the powder used for the growth. It should be noted that the initial $^{100}$MoO$_3$ was contaminated by iron at the level of 8~ppm. Besides, the special set of lab-ware used for purification of the small amount of enriched material was not perfectly clean. We hope to improve substantially the optical properties of the enriched crystals in the course of the R\&D in progress now. It is expected that the increase of the purification-cycle scale should improve the quality of the Zn$^{100}$MoO$_4$ compound. Anyway, as it will be demonstrated in the next Section, even the present optical quality of the enriched crystal scintillators is high enough to utilize the material for the development of high-performance scintillating bolometers.

The yield of the crystal boule is 83.7\%. This efficiency is unachievable in the ordinary Czochralski method (maximum $30-45\%$). It should be noted that some amount of Zn$^{100}$MoO$_4$ crystal ($\approx 1.3$ g) remained in the seed. Taking into account the mass of the residual of the melt after the growth process (30.83 g) one can estimate that the losses in the crystal growth process are about 0.6\%.

The data on losses of enriched molybdenum in all the stages of crystal
scintillator production are summarized in Table \ref{100Mo-loss}.
The amount of losses is comparable to that of enriched
cadmium in the production of cadmium tungstate crystal
scintillators from $^{106}$Cd (2.3\%) \cite{Bel10a} and $^{116}$Cd
($\approx 2\%$) \cite{Barabash:2011}. An R\&D program to decrease
the losses of enriched molybdenum by a factor $1.5-2$ by
optimization of all the stages of the production process is in
progress.

\nopagebreak

\begin{table}[htb]
\caption{Irrecoverable losses of enriched molybdenum in all
the stages of Zn$^{100}$MoO$_4$ crystal scintillator production.}
\begin{center}
\begin{tabular}{ll}
\hline
 Stage   &  Loss  \\
 \hline
 Sublimation of $^{100}$MoO$_3$             & 1.4\% \\
 Recrystallization from aqueous solutions   & 2\%   \\
 Crystal growth                             & 0.6\%  \\
\hline
 Total                                      & 4\%  \\
 \hline
\label{100Mo-loss}
\end{tabular}
\end{center}
\end{table}

\section{Fabrication and operation of two enriched Zn$^{100}$MoO$_4$ scintillating bolometers}
\label{sec:bolo}

\begin{figure}[htb]
\subfigure[]{
  \includegraphics[width=0.45\textwidth]{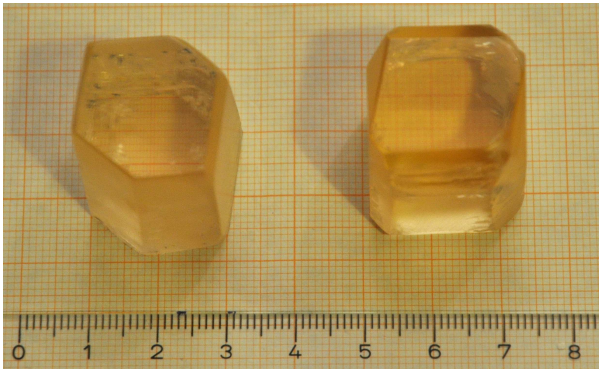}
}
\subfigure[]{
  \includegraphics[width=0.45\textwidth]{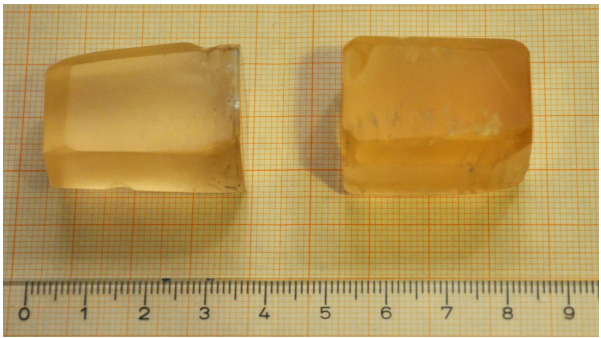}
}
\caption{Top (a) and side (b) views of the zinc molybdate single crystals obtained from material enriched in $^{100}$Mo to 99.5\%. The  Zn$^{100}$MoO$_4$-top sample (see text), at the left, has a mass of 59.2 g and the Zn$^{100}$MoO$_4$-bottom sample, at the right, of 62.9 g.}
\label{fig:cryst}
\end{figure}

Two samples with a similar size (Zn$^{100}$MoO$_4$-top and Zn$^{100}$MoO$_4$-bottom, with masses of 59.2 and 62.9 g respectively) were produced from the Zn$^{100}$MoO$_4$ boule shown in Fig.~\ref{fig:boule}. The sample Zn$^{100}$MoO$_4$-top corresponds to the top part of the boule, close to its ingot, and characterized by a less intense orange color (corresponding probably to less chemical impurities and defect concentration). A photograph of the produced crystal elements is shown in Fig.~\ref{fig:cryst}. The two samples are only approximately rectangular. Their shape is irregular as it was decided to keep their mass as high as possible in order to make the bolometric tests more significant. We took just the precaution to get two flat parallel bases in order to facilitate holding the crystals by polytetrafluoroethylene (PTFE) elements in the bolometer construction.

The two Zn$^{100}$MoO$_4$ crystals were used to assemble an array of scintillating bolometers. Each sample was equipped with a neutron trasmutation doped (NTD) Ge thermistor for the read-out of the thermal signals. The thermistors have a mass of about 50 mg, a resistance at 20 mK of $\sim 500$ k$\Omega$ and a logarithmic sensitivity $A=-d\log(R)/d\log(T) \simeq 6.5$. They were attached at the crystal surface by using six epoxy glue spots and a 25 $\mu$m thick Mylar spacer (which was removed after the gluing procedure). In addition, each crystal was provided with a heating element glued by means of one epoxy spot, consisting of a resistive meander of heavily-doped silicon with a low-mobility metallic behavior down to very low temperatures. The purpose of this heater is to provide periodically a fixed amount of thermal energy in order to control and stabilize the thermal response of the Zn$^{100}$MoO$_4$ bolometers. 

\begin{figure}
\centering
\includegraphics[width=0.45\textwidth]{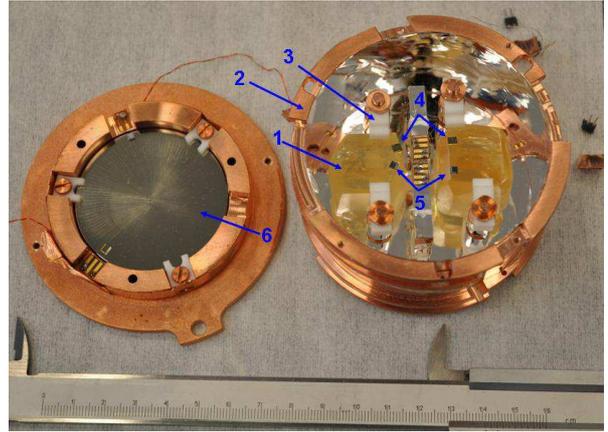}
\caption{Photograph of the assembled Zn$^{100}$MoO$_4$ bolometer array together with 
the photodetector: 
(1) Zn$^{100}$MoO$_4$ crystals with mass of 59.2 g (left) and 62.9 g (right); 
(2) copper holder of the detectors with the light reflecting foil fixed inside; 
(3) PTFE supporting elements; 
(4) NTD Ge thermistors; 
(5) heating devices; 
(6) light detector.}
\label{fig:detector}
\end{figure}

The samples were assembled inside a copper holder by using PTFE elements. A light detector (LD)~\cite{Ten12}, made of $\oslash$50$\times$0.25~mm high-purity Ge disks and instrumented with an NTD Ge thermistor was mounted above $\approx$2 mm from the top face plane of the Zn$^{100}$MoO$_4$ crystals to collect the emitted scintillation light. The inner surface of the copper holder was covered by a reflecting foil (VM2000, VM2002 by 3M) to improve light collection. The Zn$^{100}$MoO$_4$ scintillating bolometer array -- before mounting of the LD -- is shown in Fig. \ref{fig:detector}.

The array was tested at very low temperatures by using a pulse-tube cryostat housing a high-power dilution refrigerator \cite{Man14} installed at the CSNSM (Orsay, France). The cryostat is surrounded by a massive shield made out of low activity lead to minimize pile-up effects, which are particularly disturbing in aboveground measurements with slow detectors such as bolo\-meters based on NTD Ge thermistors. 

A room-temperature low-noise electronics, consisting of DC-coupled voltage-sensitive amplifiers~\cite{Arn02} and located inside a Faraday cage, was used for the read-out of the NTD Ge thermistors. Data streams were recorded by a 16 bit commercial ADC with 10 kHz sampling frequency. The test was performed at three base temperatures: 13.7 mK (over 18.3 h), 15 mK (over 4.8 h), and 19 mK (over 24.2 h). The last working point was chosen with the aim to simulate the typical temperature conditions expected in the EDELWEISS set-up \cite{Arm11}, an apparatus to search for dark matter which will be available for our next test deep underground in the Modane underground laboratory (France). 

Several experimental parameters were evaluated for each set of measurements in order to extract information about the bolometric performances of the Zn$^{100}$MoO$_4$ array. In particular, we have registered the thermistor resistance ($R_{bol}$) at the working temperature, the signal amplitude ($A_{signal}$) for a unitary deposited energy, the full width at half maximum baseline width (FWHM$_{bsl}$), and pulse rise ($\tau_R$) and decay ($\tau_D$) times. These last two parameters were computed from 10\% to 90\% and from 90\% to 30\% of the signal maximum amplitude respectively. An overview of the detector performance at 13.7 mK and 19 mK in terms of the mentioned parameters is provided in Table \ref{tab:perf}. Similar values were obtained with the Zn$^{100}$MoO$_4$ detectors cooled down to 15 mK, but these results are omitted here due to the very short duration of this measurement. 

\begin{table}[!htb]
\caption{Experimental parameters (see text) for the Zn$^{100}$MoO$_4$ scintillating bolometers array registered in aboveground measurements at 13.7 mK (first row) and 19 mK (second row). An event distribution within energy ranges of 500--3000 keV and 10--30 keV for the Zn$^{100}$MoO$_4$ bolometers and for the LD respectively were used to evaluate the parameters $\tau_R$ and $\tau_D$.}
\footnotesize
\centering
\begin{tabular}{cccccc}
\hline
Detector & $R_{bol}$  & $A_{signal}$ & FWHM$_{bsl}$ & $\tau_R$ & $\tau_D$ \\
~       & (M$\Omega$) & ($\mu$V/MeV) & (keV) & (ms) & (ms) \\
\hline
top     & 1.54 & 86.8  & 1.4(1) &  9.0 & 46.3 \\
~       & 1.17 & 65.0  & 1.8(1) &  8.9 & 48.4 \\
\hline
bottom  & 1.82 & 95.8  & 1.8(1) &  5.5 & 26.2 \\
~       & 1.35 & 84.2  & 2.4(1) &  5.8 & 30.7 \\
\hline
LD      & 0.97 & 409  & 0.28(1) &  2.5 & 14.8 \\
~       & 0.81 & 336  & 0.37(2) &  2.5 & 15.5 \\ 
\hline
\end{tabular}
   \label{tab:perf}
\end{table}

\begin{figure}
\centering
\includegraphics[width=0.45\textwidth]{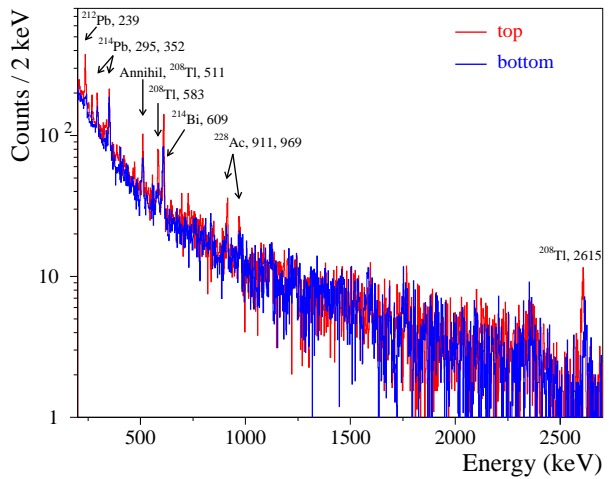}
\caption{The energy spectrum accumulated in aboveground measurements over 18.3 h 
by two Zn$^{100}$MoO$_4$ scintillating bolometers (top and bottom) mounted in one holder.  
The detector was operated at 13.7 mK and irradiated by low-active $^{232}$Th source 
and environmental $\gamma$s. The energy of $\gamma$ peaks are in keV.}
\label{fig:bg}
\end{figure}

The optimum filter procedure~\cite{Gat86}, typically used for the analysis of bolometric data, was applied 
to extract the amplitudes of each recorded signal. The calibration of the LD was performed by using a weak $^{55}$Fe source. The energy resolution (FWHM) of the LD at 5.9 keV of $^{55}$Fe was 0.42(2) keV at 13.7 mK and 0.57(4) keV at 19 mK. The energy scale of the Zn$^{100}$MoO$_4$ bolometers was determined by means of a low-activity $^{232}$Th source and $\gamma$ quanta from natural radioactivity (mainly $^{226}$Ra daughters). The energy 
spectra accumulated by the Zn$^{100}$MoO$_4$ detectors operated at 13.7 mK are shown in Fig. \ref{fig:bg}. The energy resolution for the detectors at 2614.5 keV of $^{208}$Tl was FWHM = 11(3) keV for the Zn$^{100}$MoO$_4$-top and FWHM = 15(3) keV for the Zn$^{100}$MoO$_4$-bottom in the measurements at 13.7 mK. As one can see from Fig. \ref{fig:bg}, the 2614.5 keV peaks have quite low statistics due to the small mass of the crystals and the short duration of the measurements. 
Therefore, it is reasonable to estimate the resolution of the detectors for the more intensive $\gamma$ lines presented in the spectra below 1 MeV. For example, the FWHM at 609.3 keV peak of $^{214}$Bi was measured as 5.0(5) keV and 10(1) keV for the Zn$^{100}$MoO$_4$ top and bottom, respectively. Experience with large-mass slow bolometric detectors shows that the energy resolution on the $\gamma$ lines is significantly worsened by pulse pile-up. We expect therefore that the energy resolution in an underground set-up is much closer to the FWHM baseline width and definitely better than 10~keV, as already observed in underground-operated natural ZnMoO$_4$ bolometers~\cite{Bee12a,Bee12c}. 

\begin{figure}
\centering
\includegraphics[width=0.45\textwidth]{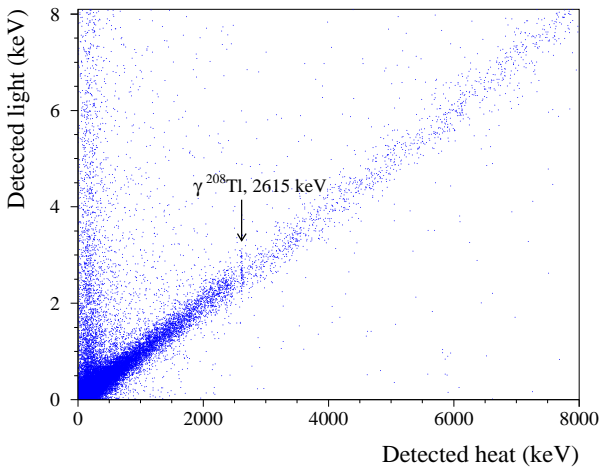}
\caption{The scatter plot of light versus detected heat based on the 13.7 mK data accumulated 
with the scintillating bolometer Zn$^{100}$MoO$_4$-top during 18.3 h of calibration measurements. 
The band populated by ${\gamma(\beta)}$ events (below 2.6 MeV) and cosmic muons is clearly visible.}
\label{fig:scatter}
\end{figure}

Plots reporting the light-to-heat signal amplitude ratio as a function of the heat signal amplitude for the data accumulated over 18.3 h in the aboveground set-up with the enriched Zn$^{100}$MoO$_4$ detectors are presented in Fig.~\ref{fig:scatter}.

\begin{figure}
\centering
\includegraphics[width=0.45\textwidth]{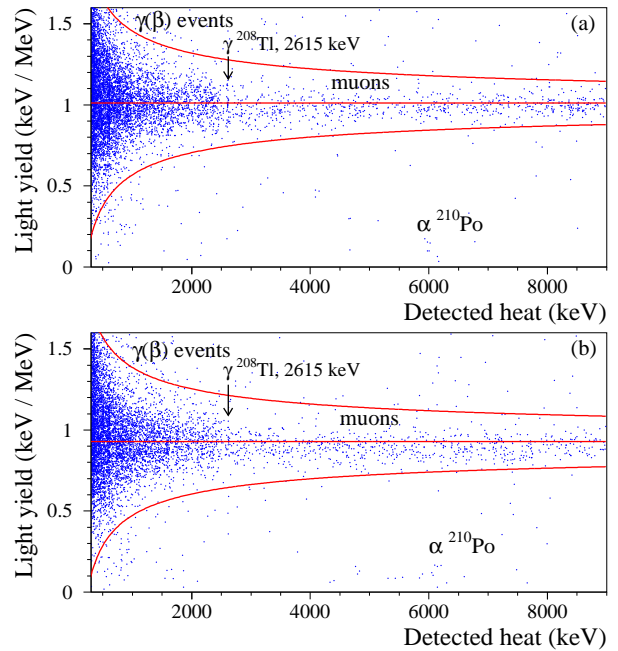}
\caption{The light yield as a function of the detected heat in aboveground measurements on scintillating bolometers based on Zn$^{100}$MoO$_4$-top (a) and Zn$^{100}$MoO$_4$-bottom (b) crystals. The data were accumulated at 13.7 mK. The events above 2.6 MeV in the ${\gamma(\beta)}$ region are caused by cosmic muons. The position of the $\alpha$ events related to $^{210}$Po is slightly shifted from the nominal value of $Q_{\alpha}$ ($\approx$ 5.4 MeV) due to a different heat response for $\alpha$ particles, already observed in this type of detectors~\cite{Bee12,Bee12a,Bee12b,Bee12c}. The mean value and three $\sigma$ intervals for LY$_{\gamma(\beta)}$ are shown by solid red lines.}
\label{fig:ly}
\end{figure}

The data allow estimating the light yield (the amount of detected light energy per particle energy measured by the deposited heat) related to ${\gamma(\beta)}$ events (LY$_{\gamma(\beta)}$). For instance, the distribution of the LY$_{\gamma(\beta)}$ versus the detected heat accumulated 
by the  Zn$^{100}$MoO$_4$-top scintillating bolometer is depicted in Fig. \ref{fig:ly}. The LY$_{\gamma(\beta)}$ was estimated by fitting the data in a 600--2700 keV interval. The fit gives similar values for all working temperatures: e.g. 1.01(11) keV/MeV (at 13.7 mK) and 1.02(11) keV/MeV (at 19 mK) for the crystal Zn$^{100}$MoO$_4$-top, and 0.93(11) keV/MeV (at 13.7 mK) and 0.99(12) \allowbreak keV/MeV (at 19 mK) for the crystal Zn$^{100}$MoO$_4$-bottom. 


%

In spite of the short duration of the measurements and the aboveground conditions, the data presented 
in Fig.~\ref{fig:ly} demonstrate an encouraging internal radiopurity level of the tested Zn$^{100}$MoO$_4$ 
crystals. There are no remarkable accumulation of counts in the region where events caused by 
$\alpha$ radionuclides from the U/Th chains are expected.\footnote{Taking into account the 
published values of quenching factors for $\alpha$ particles ($\approx$ 0.12--0.18) derived 
in measurements with ZnMoO$_4$-based bolometers~\cite{Bee12,Bee12a,Bee12b,Bee12c}, we consider all the events with energy 
in the range 4--9 MeV and light yield below 0.2 keV/MeV as related to potential $\alpha$ 
contaminants.} A weak event cluster probably related to $^{210}$Po is used to determine the activity of this radionuclide, considering a 200 keV interval around its centroid. The analysis of the data accumulated at 13.7 mK and 19 mK with the Zn$^{100}$MoO$_4$-top and Zn$^{100}$MoO$_4$-bottom bolometers gives an activity of $^{210}$Po in the crystals at the level of 1.1(3) mBq/kg and 1.2(3) mBq/kg, respectively. These activities are similar to those measured precisely in natural ZnMoO$_4$ bolometers~\cite{Bee12c}. It should be noted that the current statistics of the $^{210}$Po counts (4--6 events depending on the set of measurements) is not enough to get precise value of the activities and to distinguish a bulk contamination of $^{210}$Po (or of the progenitor $^{210}$Pb) from surface pollution. The radiopurity of the Zn$^{100}$MoO$_4$ crystals will be precisely determined in next measurements in underground conditions, but the absence of significant $\alpha$ peaks at this level indicates that the contamination of the harmful nuclides $^{228}$Th and $^{214}$Bi~\cite{Bee12}  does not exceed a level of a few mBq/kg.

Another important consideration regards the good reproducibility of the behaviour of the two detectors. The slight differences in the operational parameters of the two devices are well within the typical spread observed in this type of detectors. Both detectors, despite some difference in the optical quality, have shown practically identical bolometric and scintillation characteristics.

\section{Conclusions and prospects}
\label{sec:concl}

A zinc molybdate crystal with a mass of 171 g was produced from molybdenum
enriched in $^{100}$Mo to 99.5\% in the framework of the LUMINEU program, after a complex and effective purification method. The output of the crystal boule is 84\%, which demonstrates an important advantage of the low-thermal-gradient Czochralski technique for crystal growing. The irrecoverable losses of enriched molybdenum were found to be of the order of a few \%. The results on the scintillating bolometers fabricated with two enriched crystals obtained from the boule show that the response of these devices meets the requirements of a high-sensitivity double beta decay search and is not distinguishable from the one observed in recent measurements performed with non-enriched detectors \cite{Bee12,Bee12a,Bee12b,Bee12c,Che13,Ber14}. Encouraging, although preliminary, results were also obtained in terms of radiopurity. One can therefore expect a successful operation of scintillating bolometers based on enriched Zn$^{100}$MoO$_4$ crystals in an underground environment. This measurement is presently under preparation.

Recently, significant improvements in the growth technology developed at NIIC (Novosibirsk, Russia) have enabled the synthesis of large regular-shape cylindrical ($\oslash$5$\times$4~cm) natural ZnMoO$_4$ crystals. By using an amount of enriched material larger than that employed in the experiment here described,\footnote{Several kilograms of enriched molybdenum, belonging partly to the Institute of Theoretical and Experimental Physics (Moscow, Russia) and partly to the Institute for Nuclear Research (Kyiv, Ukraine) are available for the development of Zn$^{100}$MoO$_4$ scintillating bolometers.} we foresee to develop Zn$^{100}$MoO$_4$ single detectors with masses in the 300 -- 500 g range in the near future. The size of these devices corresponds to the one envisaged for the single module of large arrays of Zn$^{100}$MoO$_4$ detectors to search for $0 \nu 2 \beta$ decay of $^{100}$Mo~\cite{Bee12}. The currently available enriched material will allow developing an array with a total mass of $\approx 15-20$~kg, corresponding to a sensitivity to the effective Majorana mass in the range 0.05 -- 0.15 eV~\cite{Bee12} and capable therefore to approach the inverted hierarchy region of the neutrino mass pattern.

\section{Acknowledgments}

The development of Zn$^{100}$MoO$_4$ scintillating bolometers is an essential part of the LUMINEU program, receiving funds from the Agence Nationale de la Recherche (France). The work was supported partly by the project ``Cryogenic detectors to search for neutrinoless double beta decay of molybdenum'' in the framework of the Programme ``Dnipro'' based on Ukraine-France Agreement on Cultural, Scientific and Technological Cooperation. The aboveground facility for the test of the scintillating bolometers was realised with the indispensable support of ISOTTA, a project funded by the ASPERA 2nd Common Call dedicated to R\&D activities. The group from the Institute for Nuclear Research (Kyiv, Ukraine) was supported in part by the Space Research Program of the National Academy of Sciences of Ukraine.

It is a pleasure to thank Stefano Nisi from the Gran Sasso laboratory (Italy)  for his kind help in the measurements of enriched molybdenum oxide by inductively coupled plasma mass-spectrometry.

\end{document}